\documentclass[english,reprint, citeautoscript, aps, prb, nobibnotes, superscriptaddress]{revtex4-1}
\usepackage[T1]{fontenc}
\usepackage[utf8]{inputenc}
\usepackage{verbatim}
\usepackage{amsmath}
\usepackage{amssymb}
\usepackage{amsfonts}
\usepackage{amstext}
\usepackage{graphicx}
\usepackage{float}

\usepackage{soul}
\usepackage{color}

\definecolor{purple}{rgb}{0.5,0,0.5}

\newcommand\Tstrut{\rule{0pt}{2.6ex}}         

\newcommand{\unit}[2]{${#1}\,\mathrm{#2}$}
\newcommand{\unitm}[2]{{#1}\,\mathrm{#2}}
\newcommand{\angstrom}{\textup{\AA}} 



\usepackage{babel}
\begin{document}

\title{Automated discovery of a robust interatomic potential for aluminum}

\author{Justin S. Smith}
\email{just@lanl.gov}
\affiliation{Theoretical Division, Los Alamos National Laboratory, Los Alamos, NM 87545, USA}
\affiliation{Center for Nonlinear Studies, Los Alamos National Laboratory, Los Alamos, NM 87545, USA}

\author{Benjamin Nebgen}
\email{bnebgen@lanl.gov}
\affiliation{Theoretical Division, Los Alamos National Laboratory, Los Alamos, NM 87545, USA}

\author{Nithin Mathew}
\affiliation{Theoretical Division, Los Alamos National Laboratory, Los Alamos, NM 87545, USA}
\affiliation{Center for Nonlinear Studies, Los Alamos National Laboratory, Los Alamos, NM 87545, USA}

\author{Jie Chen}
\affiliation{Materials Division, Los Alamos National Laboratory, Los Alamos, NM 87545, USA}

\author{Nicholas Lubbers}
\affiliation{Computer, Computational, and Statistical Sciences Division, Los Alamos National Laboratory, Los
Alamos, NM 87545, USA}

\author{Leonid Burakovsky}
\affiliation{Theoretical Division, Los Alamos National Laboratory, Los Alamos, NM 87545, USA}

\author{Sergei Tretiak}
\affiliation{Theoretical Division, Los Alamos National Laboratory, Los
Alamos, NM 87545, USA}

\author{Hai Ah Nam}
\affiliation{Computer, Computational, and Statistical Sciences Division, Los Alamos National Laboratory, Los
Alamos, NM 87545, USA}

\author{Timothy Germann}
\affiliation{Theoretical Division, Los Alamos National Laboratory, Los
Alamos, NM 87545, USA}

\author{Saryu Fensin}
\affiliation{Materials Division, Los Alamos National Laboratory, Los Alamos, NM 87545, USA}

\author{Kipton Barros}
\email{kbarros@lanl.gov}
\affiliation{Theoretical Division, Los Alamos National Laboratory, Los
Alamos, NM 87545, USA}

\begin{abstract}
Accuracy of molecular dynamics simulations depends crucially on the interatomic potential used to generate forces. The gold standard would be first-principles quantum mechanics (QM) calculations, but these become prohibitively expensive at large simulation scales. Machine learning (ML) based potentials aim for faithful emulation of QM at drastically reduced computational cost. The accuracy and robustness of an ML potential is primarily limited by the quality and diversity of the training dataset. Using the principles of active learning (AL), we present a highly automated approach to dataset construction. The strategy is to use the ML potential under development to sample new atomic configurations and, whenever a configuration is reached for which the ML uncertainty is sufficiently large, collect new QM data. Here, we seek to push the limits of automation, removing as much expert knowledge from the AL process as possible. All sampling is performed using MD simulations starting from an initially disordered configuration, and undergoing non-equilibrium dynamics as driven by time-varying applied temperatures. We demonstrate this approach by building an ML potential for aluminum (ANI-Al). After many AL iterations, ANI-Al teaches itself to predict properties like the radial distribution function in melt, liquid-solid coexistence curve, and crystal properties such as defect energies and barriers. To demonstrate transferability, we perform a 1.3M atom shock simulation, and show that ANI-Al predictions agree very well with DFT calculations on local atomic environments sampled from the nonequilibrium dynamics. Interestingly, the configurations appearing in shock appear to have been well sampled in the AL training dataset, in a way that we illustrate visually.
\end{abstract}

\maketitle

\section{Introduction}

Given sufficient training data, ML models show great promise to accelerate scientific simulation, e.g., by emulating expensive computations at high accuracy but much reduced computational cost. ML modeling of atomic scale physics is a particularly exciting area of development~\cite{Rupp2012,Bleiziffer2018MachineCalculations,Butler2018MachineScience,Sifain2018}. Provided sufficient training data, ML models suggest the possibility for development of models with unprecedented transferability. Applications to materials physics, chemistry, and biology are innumerable. To give some examples, simulations for crystal structure prediction, drug development, materials aging, and high strain/strain-rate deformation would all benefit from better interatomic potentials.
  
Machine learning (ML) of interatomic potentials is a rapidly advancing topic, for both materials physics~\cite{Behler2007,Handley2014,Szlachta2014,Li2015,Thompson2015SpectralPotentials,Kruglov2017,Huan2017,Jindal2017,Botu2017MachineOutlook,Schutt2018,Deringer2018ComputationalTheory,Suwa2019MachineElectrons,Pozdnyakov2019FastPotential,Liu2020StructurePotential} and chemistry~\cite{Morawietz2012ACharges,Smith2017,Yao2017ThePhysics,Lubbers2018,Nguyen2018ComparisonExpansions,Gastegger2017MachineSpectra,Smith2019ApproachingLearning,Schran2019AutomatedGround}. Training datasets are calculated from computationally expensive \textit{ab initio} quantum mechanics methods, most commonly density functional theory (DFT)~\cite{Engel2011DensityTheory}. Trained on this data, an ML model can be very successful in predicting energy and forces for new atomic configurations. ML potentials typically assume very little beyond symmetry constraints (e.g., translation and rotation invariance) and spatial locality (each atomic force only depends on neighboring atoms within a fixed radius, typically of order 5 to \unit{10}{\angstrom}). Long-range Coulomb interactions or dispersion corrections may also be added~\cite{Yao2017ThePhysics,Unke2019PhysNet:Charges}.

 For large-scale molecular dynamics (MD) simulations, so-called classical potentials are usually the tool of choice. Such potentials are relative simple and computationally efficient. Although effective for many purposes, classical potentials may struggle to achieve broad transferability. For example, it is not easy to design a single classical potential that correctly describes multiple incompatible crystal phases and the transitions between them. Consequently, assumed functional forms for classical potentials tend to grow more flexible over time. For example, the embedded atom method (EAM)~\cite{Daw1983SemiempiricalMetals}, has lead to generalizations such as modified EAM (MEAM)~\cite{Lee2010TheSimulations} and multistate MEAM~\cite{Baskes2007MultistateMethod}.
  
 In contrast to classical potentials, the ML philosophy is to begin with a functional form of utmost flexibility. For example, a modern neural network-based ML potential may contain $\sim 10^5$ fitting parameters. If properly trained, recent work suggests that the accuracy of ML potentials can approach that of the underlying \textit{ab initio} theory (e.g. DFT or coupled cluster)~\cite{Smith2017,Yao2017ThePhysics,Kobayashi2017NeuralAlloys,Sifain2018,Zhang2019ActiveSimulation,Zhang2019DP-GEN:Models,Pun2019PhysicallyinformedMaterials, Smith2019ApproachingLearning}. Although slower than classical potentials, ML potentials are vastly faster than, say, reference DFT calculations. The main limitation on the accuracy and transferability of an ML potential is the quality and broadness of the training dataset.
 
 In this paper, we design an active learning approach for automated dataset construction suitable for materials physics, and demonstrate its power by building a robust potential for aluminum that we call ANI-Al. Distinct from previous works, here \textit{the active learning scheme receives practically no expert guidance}. In particular, we do not seed the training dataset with any crystal or defect structures; the active learner begins only with fully randomized atomic configurations. By leaving the search space of possibly relevant atomic configurations unspecified, we aim to build a model that is maximally general. If successful, the model should remain accurate when presented with complex atomic configurations that may arise in a variety of highly non-equilibrium dynamics.
 
  The basic steps of active learning (AL) for atomic-scale modeling are to sample new atomic configurations, query the ML model for uncertainty in its predictions, and selectively collect new training data that would best improve the model~\cite{Reker2015Active-learningDiscovery,Podryabinkin2017ActivePotentials,Gastegger2017MachineSpectra,Gubaev2018MachineLearning,Bernstein2019DeSurfaces,Jinnouchi2020On-the-FlySimulations,Sivaraman2020Machine-learnedDioxide}. Previous work employed AL to drive nonequilibrium sampling of large datasets through organic chemical space, yielding the highly general ANI-1x potential~\cite{Smith2018LessLearning}. Other recent research by Gubaev et al.~\cite{Gubaev2019AcceleratingPotentials} has explored the use of AL with moment tensor potentials to construct atomistic potentials for materials. Zhang et al. also applied AL to materials using the deep potential model~\cite{Zhang2019ActiveSimulation} for MgAl alloys. AL was used by Deringer, Pickard, and Csányi to build an accurate and general model for elemental Boron~\cite{Deringer2018Data-DrivenBoron}.
 
  The AL procedure developed in the present work will be discussed in detail below, but briefly, there is a loop over three main steps: (1) Using the best ANI-Al models available, MD simulations with time-varying temperatures are performed to sample new atomic configurations; (2) an ML uncertainty measure determines whether the sampled configurations would be useful for inclusion in the training data and, if so, new DFT calculations are run; and (3) new ANI-Al models are trained with all available training data. The starting point for AL is an initial training dataset consisting of DFT calculations on randomized (disordered) atomic configurations. Each MD sampling trajectory is also initialized to a random disordered configuration, with random density. Crucially, the AL scheme receives no \textit{a priori} guidance about the relevant configuration space it should sample. Nonetheless, after enough iterations, the AL procedure eventually encounters configurations that locally capture characteristics of crystals such as FCC, HCP, BCC, and many others. The AL algorithm is readily parallelizable; we employed hundreds of nodes on the Sierra supercomputer to collect the final ANI-Al dataset consisting of about 6,000 DFT calculations.
  
  We demonstrate below that ANI-Al does a good job in predicting standard properties of the stable FCC crystal and liquid phases for aluminum. It also effectively predicts defect energy barriers and the liquid-solid coexistence curve. We emphasize, however, that our AL procedure is designed to primarily sample \textit{disordered and partially-ordered configurations}. Certainly many locally crystal-like configurations will appear through the dynamics, but these will typically be imperfect.
  
  Our motivation for designing an unconstrained sampling strategy was to maximize ANI-Al accuracy when applied to extreme and highly nonequilibrium processes. As a test, we performed a 1.3M atom shock simulation, and verified the ANI-Al predicted forces by performing new DFT calculations on randomly sampled local atomic environments. For this simulation, ANI-Al force prediction errors (per component) were of order \unit{0.03}{eV/\angstrom}, whereas typical force magnitudes ranged from 1 to \unit{2.5}{eV/\angstrom}. In terms of absolute force accuracy, ANI-Al performs nearly as well for extreme shock simulations as it does for \textit{equilibrium} crystal or liquid simulations.
  
  Finally, we will present a visualization of the high-dimensional space of all sampled atomic configurations by embedding them into an abstract, two-dimensional space. The results indicate that AL does a good job of sampling many different crystal symmetries, the liquid phase, and the highly defected configurations that appear in shock. Interestingly, these three types of data are observed to be largely non-overlapping in the 2D embedding space.
 
   
 \section{Building the ANI-A\lowercase{l} potential} \label{Methods}
 
 This section presents details of the automated procedure to build ANI-Al, our general purpose machine learning potential for bulk aluminum.

  \subsection{The ANI machine learning model}  \label{meth:animodel}
    ANI is a neural network architecture for modeling interatomic potentials. Our prior work with ANI has largely focused on modeling clusters of organic molecules~\cite{Smith2019ApproachingLearning}. A variety of ANI potentials are available online~\cite{2017NeuroChemgithub.com/atomistic-ml/neurochem}. Here we present ANI-Al, our ANI model for aluminum in both crystal and melt phases.
    
    Our training data consists of DFT calculations, evaluated on ``interesting'' atomic configurations, as identified by the active learning procedure (Sec.~\ref{meth:activelearn}). We used PBE functional, with parameters described in Sec. 1.1 of the SI. One point to mention is that our $3\times 3\times 3$ $k$-space mesh was, in retrospect, perhaps too small. For the varying box sizes of our training data, this corresponds to 31.5 to 51 $k$-points per $\angstrom^{-1}$. A more careful choice would be 57 $k$-points per $\angstrom^{-1}$ independent of system size~\cite{Botu2015AdaptiveDynamics}.
    
    The input to ANI is an atomic configuration (nuclei positions and species). To describe these configurations in a rotation and translation invariant way, ANI employs Behler and Parrinello~\cite{Behler2007} type atomic descriptors, but with modified angular symmetry functions~\cite{Smith2017}. Details of all model hyperparameters are provided in Sec.~1.3 of the SI. The most important hyperparameter is the \unit{7}{\angstrom} interaction cutoff distance, which we selected based on careful trial and error. Other hyperparameters, such as the number of symmetry functions, were largely reused from previous studies~\cite{Smith2019ApproachingLearning}. ANI's total energy prediction is computed as a sum over local contributions, evaluated independently at each atom. Each local energy contribution is calculated using knowledge of all atoms within the \unit{7}{\angstrom} cutoff. Using backpropagation, one can efficiently calculate all forces as gradients of the predicted energy.
    
   Each DFT calculation outputs the total system energy $E$ and the forces $\mathbf{f}_{j} = \partial E / \partial \mathbf{r}_j$ for all atoms $j=1\dots N$. Our loss function for a single DFT calculation,
    \begin{equation}
        L \propto \left(\hat{E}-E\right)^2+\ell_0^2 \sum_{j=1}^{N}{\left(\hat{\mathbf f}_{j}-\mathbf f_{j}\right)^2},
    \end{equation}
    is a measure of disagreement between the ANI predictions for energy, $\hat{E}$, and forces, $\hat{\mathbf f}_{j} = \partial\hat{E} / \partial \mathbf{r}_j$, and the DFT reference data.  A length hyperparameter $\ell_0$ is empirically selected so that energy and force terms have comparable magnitude. In our tests, the specific choice of $\ell_0$ did not strongly affect the quality of the final model.
    
    Training ANI corresponds to tuning all model parameters to minimize the above loss, summed over all DFT calculations in the dataset. For stochastic gradient descent, each training iteration requires estimating the $\partial L / \partial W_i$ for all model parameters $W_i$ (in our case, there are order $10^5$ parameters). Because forces $\hat{\mathbf f}_{j}$ appear in $L$, calculating $\partial L / \partial W_i$ seems to involve all \emph{second} derivatives of the ANI energy output, i.e., $\partial^2 \hat E / \partial W_i \partial \mathbf r_j$. Fortunately, direct calculation of these can be avoided. Instead, we employ the recently proposed method of Ref.~\onlinecite{Smith2020SimpleData} to efficiently calculate all $\partial L / \partial W_i$ in the context of our C++ Neurochem implementation of ANI~\cite{2017NeuroChemgithub.com/atomistic-ml/neurochem}. A brief summary of the method is presented in Sec.~1.5 of the SI. With this method, the total cost to calculate all $\partial L / \partial W_i$ is within a factor of two of the cost to calculate all forces.

    To improve the quality of our predictions, a single ANI-Al model actually employs ensemble-averaging over 8 neural networks. Each neural network in the ensemble is trained to the same data, but using an independent random initialization of the model parameters. We observe that ensemble-averaged energy and force errors can be up to 20\% and 40\% smaller, respectively, than those of a single neural network prediction. 

  \subsection{Active learning} \label{meth:activelearn}
  \begin{figure}
   \centering
    \includegraphics[width=1.0\columnwidth]{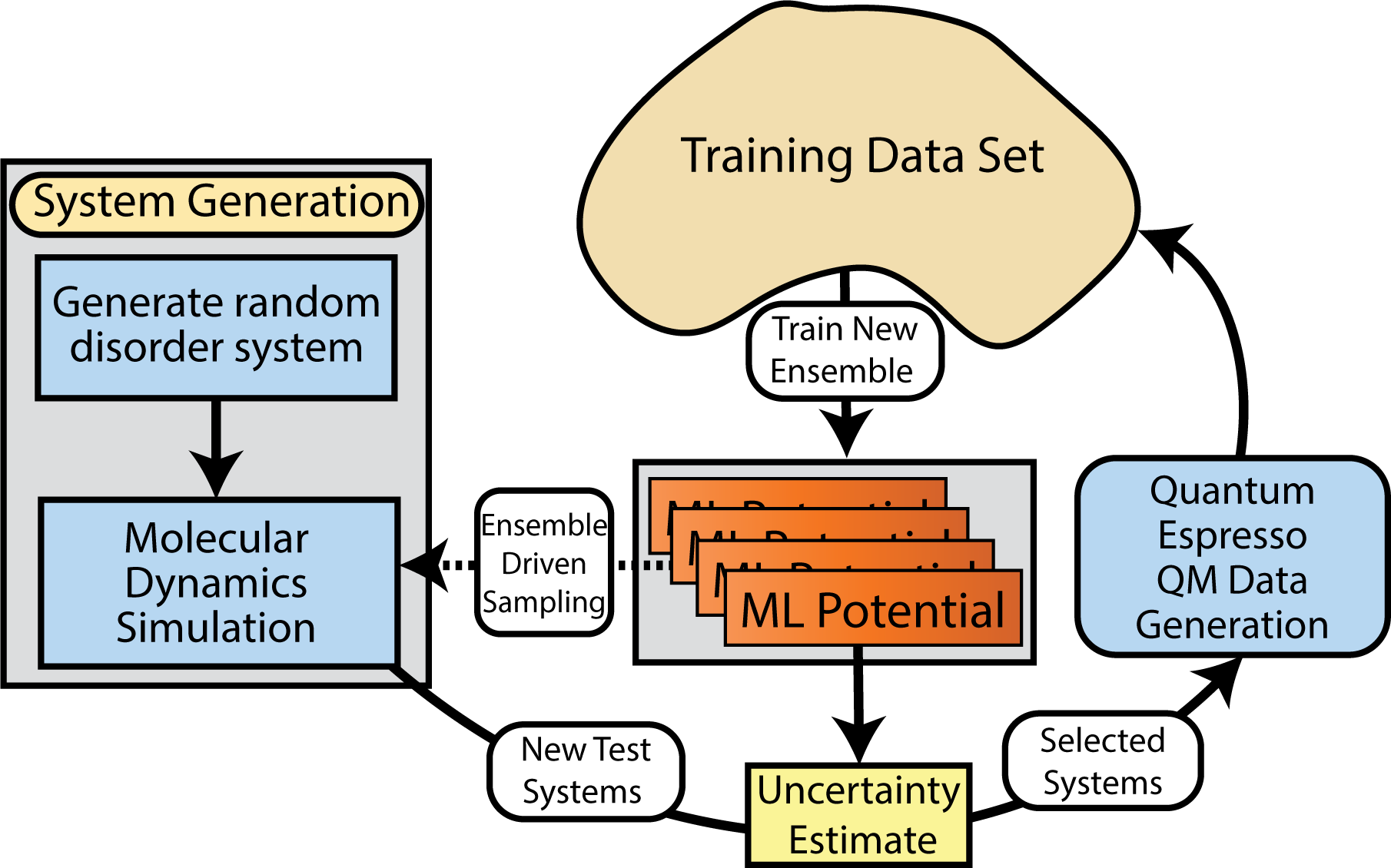}
    \caption{Diagram of the active learning sampling algorithm employed in this work. Multiple such cycles can be run simultaneously, with occasional synchronization points to combine all new data into a single global dataset. The MD sampling, DFT data generation, and ML model training all benefit from GPU-acceleration.}
    \label{fig:aldatagen}
  \end{figure}
  
  \subsubsection{Overview}

  The active learning process employed here is similar to that in previous work,~\cite{Smith2018LessLearning} adapted for materials problems and efficient parallel execution on hundreds to thousands of nodes on the Sierra supercomputer. We first train an initial model to a dataset of about 400 random disordered atomic configurations, generated as in Sec.~\ref{meth:sampling:randomdisorder}. Next, we begin the AL procedure. Using the current ML potential, we simulate many MD trajectories, each initialized to a random disordered configuration. During these simulations the temperature is dynamically varied to diversify the sampled configurations (Sec.~\ref{meth:activelearn:tempschedule}). As these MD simulations run, we look at the variance of the predictions for the 8 neural networks within an ensemble to determine whether the model is operating as expected~\cite{Seung1992QueryCommittee}. Prior work indicates that this measured ensemble variance correlates reasonably well with actual model error~\cite{Smith2018LessLearning}. If the ensemble variance exceeds a threshold value, then it seems likely that collecting more data would be useful to the model. In this case, MD trajectory is terminated and the final atomic configuration is placed on a queue for DFT calculation and addition to the training dataset. Periodically, the ML model is retrained to the updated training dataset. This AL loop is iterated until the cost of the MD simulations becomes prohibitively expensive. Specifically, we terminate the procedure when typical MD trajectories reach about \unit{250}{ps} (about $2.5\times 10^5$ timesteps) without uncovering any weaknesses in the ML model. The final active learned dataset contains 6,352 DFT calculations, each containing 55 to 250 atoms, and having varying levels of disorder.
  
  We emphasize that this active learning procedure is fully automated, and receives no direct guidance regarding atomic configurations of likely relevance, such as crystal structures. The initial training dataset consists only of disordered atomic configurations, and every MD simulation is initialized to a disordered configuration.  The MD simulations use only forces as predicted by the most recently trained ML potential. After many active learning iterations, the MD simulations will hopefully be sufficiently robust to support nucleation into, e.g., the crystal ground state, and then the active learning scheme can begin to collect this type of training data. In this sense, the active learning scheme must \emph{automatically discover} the important low energy and non-equilibrium physics.

  Section~1.2 of the SI gives further details regarding the active learning procedure. 
  
  \subsubsection{Randomized atomic configurations} \label{meth:sampling:randomdisorder}
  We employ randomized atomic configurations to collect an initial dataset of DFT calculations, and to initialize all MD simulations for AL sampling. The procedure to randomize a supercell is as follows:
  \begin{enumerate}
      \item Randomly sample each of the three linear dimensions of the orthorhombic supercell uniformly from the range 10.5 to \unit{17.0}{\angstrom}
      \item Randomly select a target atomic density $\rho$ uniformly from the range 1.80 to \unit{4.05}{g/cm^3}.
      \item Iteratively place atoms randomly in the supercell. If the proposed new atom lies within a distance $r_\textrm{min} = \unitm{1.8}{\angstrom}$ of an existing atom (i.e., roughly the van der Waals radius), that placement is rejected as unphysical. Placement of atoms is repeated until the target density $\rho$ has been reached.
  \end{enumerate}
  
  \subsubsection{Nonequilibrium temperature schedule} \label{meth:activelearn:tempschedule}
  To maximize the diversity in active learning sampling, we perform the MD simulations with a Langevin thermostat using a temperature that varies in time according to a randomized schedule. Compared with previous work that sampled from a specific temperature quench schedule~\cite{Deringer2017MachineCarbon}, here we employ a more diverse and randomly generated collection of temperature schedules.
   
  Starting at time $t=0$, and running until $t=t_\textrm{max} = \unitm{250}{fs}$, the applied temperature is,
  \begin{equation}\label{eq:Tsch}
      T(t) = T_\textrm{start} + \frac{t}{t_\textrm{max}} (T_\textrm{end} - T_\textrm{start}) +  T_\textrm{mod} \sin^2 (\pi t / t_0) \\
  \end{equation}
  The first two terms linearly ramp the background temperature. The initial temperature $T_\textrm{start}$ is randomly sampled from the range \unit{10}{K} to \unit{1000}{K}. The final background temperature $T_\textrm{end}$ is randomly sampled from the range \unit{10}{K} to \unit{600}{K}. The third term in Eq.~\eqref{eq:Tsch} superimposes temperature oscillations. The modulation scale $T_\textrm{mod}$ is randomly sampled from the range \unit{0}{K} to \unit{2000}{K}. The oscillation period $t_0$ is randomly sampled from the range \unit{10}{ps} to \unit{50}{ps}.
  
  By spawning MD simulations with many different temperature schedules, we hope to observe a wide variety of nonequilibrium processes. Given that each MD simulation begins from a disordered melt configuration, we hope that the nonequilibrium dynamics will automatically produce: (1) nucleation into various crystal structures (in particular, the ground-state FCC crystal), (2) a variety of defect structures and dynamics (dislocation glide, vacancy diffusion, etc.) and (3) rapid quenches into disordered glass phases. Acquiring snapshots from these types of dynamics will be crucial to the diversity of the training dataset and, thus, to the overall generality of the ANI-Al potential.

\section{Accuracy Benchmarks}

Here we present a variety of benchmarks for ANI-Al, our machine learned potential for bulk aluminum. As described in Sec.~\ref{Methods}, ANI-Al is trained from over 6,000 DFT calculations that were carefully selected using an iterative ``active learning'' procedure.

   \begin{figure}
    \includegraphics[width=\linewidth]{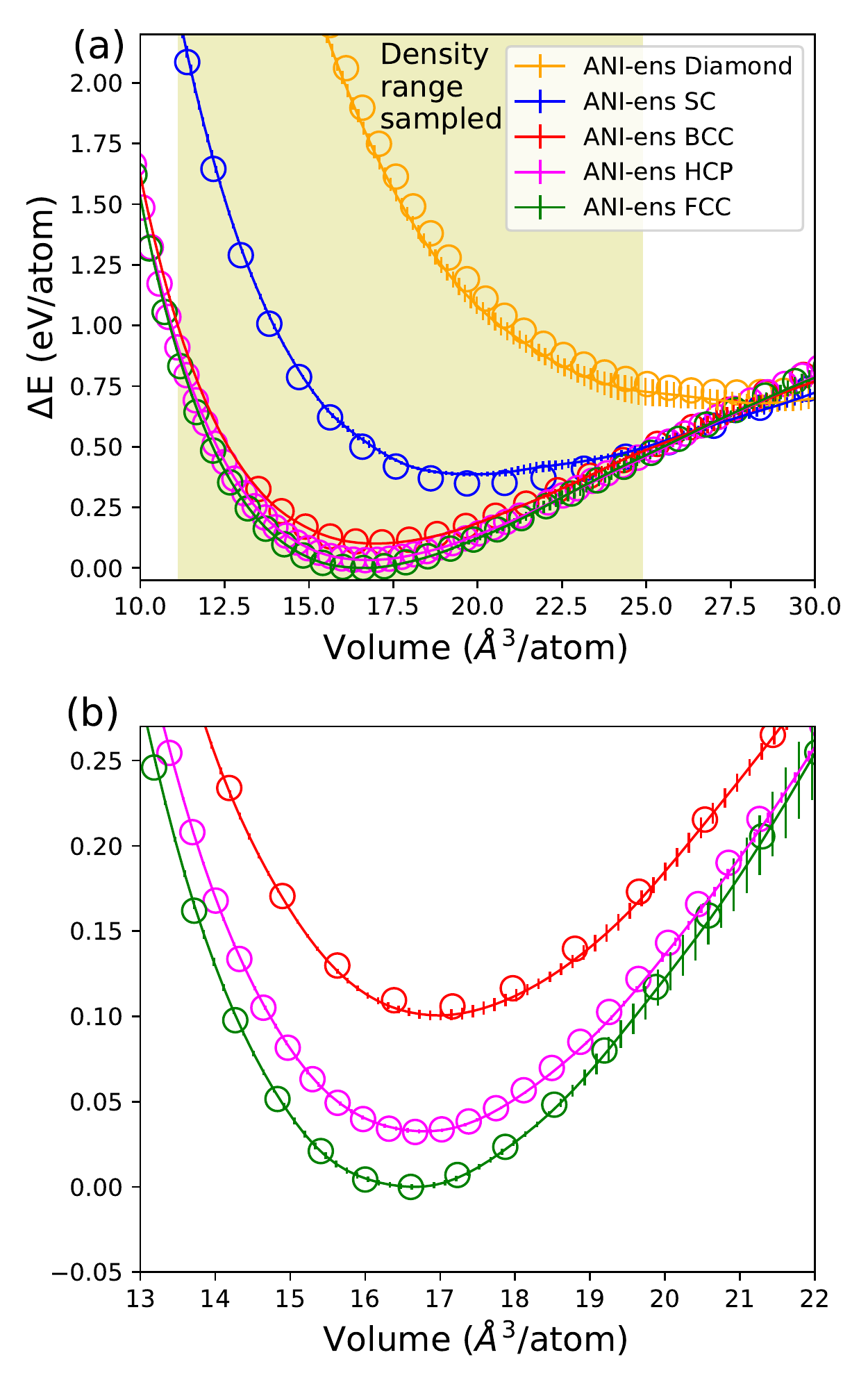}
    \caption{Crystal energies (per atom) as a function of volume (per atom), relative to the ground state. Solid lines represent ANI-Al predictions and circles represent DFT reference calculations. Vertical bars represent sample variance of the eight neural networks comprising the (ensembled) ANI-Al model. Panel (b) is a magnification of panel (a) near the energy minima. The highlighted yellow region (about 11 to \unit{25}{\angstrom^3/atom}) indicates the approximate range of densities sampled in the training data. Crystal structures are diamond, simple cubic (SC), body centered cubic (BCC), hexagonal close packed (HCP), and face centered cubic (FCC).}
    \label{fig:coldcurveAl}
  \end{figure}
  
  \subsection{Predicting crystal energies}
  Figure~\ref{fig:coldcurveAl} shows ANI-Al predicted energies (solid lines) for select crystal structures. ANI-Al correctly predicts that FCC has the lowest energy of all crystals considered; more crystal energies are compared in Table~S5 of the SI. Vertical bars show the sample variance over the eight neural networks that comprise a single ANI-Al model (i.e., the uncertainty measure used within the active learning procedure). DFT reference data is shown in circles.
  
  For both ANI-Al and DFT calculations, energies are measured relative to the FCC ground state. Let $\epsilon_x$ represent the error of the ANI-Al predicted energy for crystal structure $x$ at its energy minimizing volume (volume is independently optimized for ANI-Al and DFT). By definition, the energy shifts are such that $\epsilon_\textrm{fcc} = 0$. After FCC, the second lowest energy structure shown in this plot is HCP, for which the ANI-Al error is $\epsilon_\textrm{hcp} = 0.42\,\textrm{meV}/\textrm{atom}$. Note that FCC and HCP are competing close-packed structures, and both can reasonably be expected to emerge in our active learning dynamics (FCC with a stacking fault looks locally like HCP). BCC, by contrast, is only physical in aluminum at much higher densities, far beyond the range of our active learning sampling. It is not surprising, therefore, that the ANI-Al error for BCC is an order of magnitude larger, $\epsilon_\textrm{bcc} = 5.3\,\textrm{meV}/\textrm{atom}$. Simple cubic and diamond crystals are less physical still, and we observe $\epsilon_\textrm{sc} = 37\,\textrm{meV}/\textrm{atom}$ and $\epsilon_\textrm{diamond} = - 44\,\textrm{meV}/\textrm{atom}$. Nonetheless, the qualitative agreement between ANI-Al and DFT observed in Fig.~\ref{fig:coldcurveAl}, even for very unphysical crystals, seems remarkable. Similar observations were made in Ref.~\onlinecite{Zhang2019ActiveSimulation}. We emphasize that in the present work, the training data includes no hand-selected crystals. Instead, all atomic configurations in the training dataset were generated using MD sampling, starting only from disordered configurations.
  
  ANI-Al predictions are most reliable for the range of densities sampled in the training data (Fig.~\ref{fig:coldcurveAl}a, yellow region). Further extrapolation of these cold curves is shown in Fig. S9 of the SI.
  
 \subsection{Predicting elastic constants}
 
 We can compare ANI-Al predicted elastic constants against experimental data. A particularly important one is the bulk modulus, which corresponds to the curvature of the FCC cold curve at its minimum (Fig.~\ref{fig:coldcurveAl}b). Experimentally, the FCC bulk modulus is measured to be \unit{79}{GPa}~\cite{Simmons1971SingleHandbook}, whereas the ANI-Al prediction is \unit{77.3}{GPa}. The full set of FCC elastic constants is measured experimentally to be, $C_{11}=\unitm{114}{GPa}$, $C_{12}=\unitm{61.9}{GPa}$, and $C_{44}=\unitm{31.6}{GPa}$~\cite{Simmons1971SingleHandbook}. For our DFT calculations, $C_{11}=\unitm{106}{GPa}$, $C_{12}=\unitm{62.3}{GPa}$, and $C_{44}=\unitm{31.6}{GPa}$. For ANI-Al, we predict $C_{11}=\unitm{117}{GPa}$, $C_{12}=\unitm{57.2}{GPa}$, and $C_{44}=\unitm{30.4}{GPa}$.
 
 The largest discrepancies between ANI-Al and DFT are observed for the elastic constants $C_{11}$ and $C_{12}$, with relative errors of 10.38\% and -8.19\%, respectively. Interestingly, the effects of these two discrepancies seems to cancel in the bulk modulus, $B = (1/3) (C_{11} + 2 C_{12})$, for which the error relative to DFT is just 0.78\%. We suspect the cancellation is not coincidence, because a similar phenomenon was observed in previous ML potentials developed for aluminum~\cite{Zhang2019ActiveSimulation, Pun2019PhysicallyinformedMaterials} (cf. Table S4 of the SI).  Elastic constants measure the response of stress to a small applied strain. Note  that stress can be inferred from forces. Therefore, for an ML model to precisely capture $C_{ij}$, its training data should ideally contain many locally-perfect FCC configurations for a variety of small strains. The mechanisms by which our active learning sampler can generate strained FCC are somewhat limited (e.g., nucleation of imperfect crystals). Future work might employ time-varying applied strains to the entire supercell, in addition to the time-varying temperatures employed in the present study.
 
 In predicting elastic constants, ANI-Al accuracy is on par with many classical potentials and existing ML potentials, as shown in Tables~S3 and S4 in the SI.
 Whereas classical potentials are usually designed to reproduce experimental elastic constants, in ANI-Al this capability is an \emph{emergent} property. Our active learning sampling discovers the FCC lattice and its properties in an automated way.

  \begin{figure}
    \includegraphics[width=1.0\columnwidth]{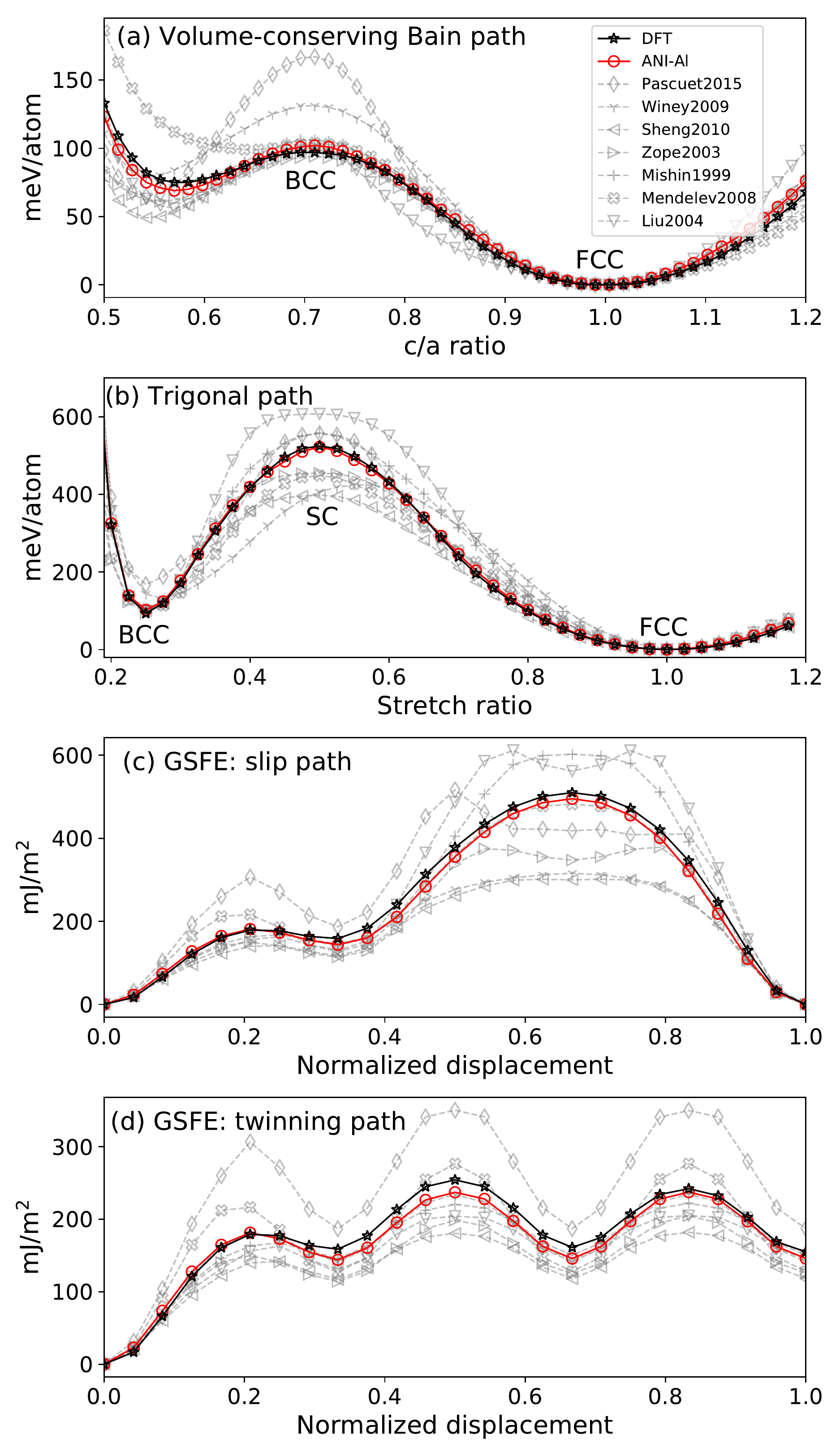}
    \caption{Transformational energy barriers. We compare ANI-Al and various classical potentials to reference DFT data. (a) Volume-conserving Bain path energies. (b) Trigonal path energies. (c) Generalized Stacking Fault Energy (GSFE) slip path. (d) GSFE twinning path.}
    \label{fig:bain_sf}
  \end{figure}
  
  \subsection{Predicting crystal energy barriers}
  
   The Bain path (Figure 3a) represents a volume-preserving homogeneous deformation that transforms between FCC and BCC crystals. Starting from the initial FCC cell ($c/a$= 1), we compress along one of the $\langle 100 \rangle$ directions (length $c$) while expanding equally in the two orthogonal directions (lengths $a=b$). The special value of $c/a=$ 1/$\sqrt{2} \approx 0.71$ would correspond to BCC symmetry. Figure~\ref{fig:bain_sf}a shows energies along this Bain path, in which $c/a$ varies continuously while conserving volume, $a^2 c$. The observed maximum at $c/a=$ 1/$\sqrt{2}$ indicates that the BCC structure is unstable to tetragonal deformation. We compare ANI-Al to DFT reference calculations, as well as seven EAM-based potentials~\cite{Mendelev2008AnalysisCu,Sjostrom2016MultiphaseTheory,Liu2004AluminiumEnergy,Mishin1999InteratomicCalculations,Zope2003InteratomicSystem,Winey2009AAluminum,Pascuet2015AtomicAlloy,Sheng2011HighlyMetals}. Figure~S2 of the SI quantifies the errors for each potential, averaged over the strain path.
   
   Figure~\ref{fig:bain_sf}b shows the energies along the trigonal deformation path, where the ideal FCC phase is compressed along the $z = \langle 111 \rangle$ crystallographic direction, and elongated equally in the two orthogonal directions $x$ and $y$, such that the total volume is conserved. We define a characteristic ``stretch ratio'' as $(L_z' /L_x')/(L_z/L_x)$ where $L_z$ and $L_x$ are the dimensions of the reference FCC simulation cell along $z$ and $x$ directions, respectively, and $L_z'$ and $L_x'$ are the dimensions of the deformed simulation cell. Stretch ratios of 1.0, 0.5, and 0.25 result in FCC, SC, and BCC phases, respectively. Good agreement is found between ANI-Al and DFT reference calculations. It can be seen from Figure~\ref{fig:bain_sf}b that SC, but not BCC, is unstable to trigonal deformation.
   
   A stacking fault in FCC represents a planar defect in which the crystal locally is in HCP configuration within the nearest neighbor shell (note that FCC and HCP are competing close packed structures). The generalized stacking fault energy (GSFE) slip path provides an estimate of the resistance for dislocation slip and the energy per unit area required to form a single stacking fault. The GSFE twinning path (also known as the generalized planar fault energy) is an extension of the slip path and provides an estimate of the energy per unit area required to form $n$-layer faults (twins) by shearing $n$ successive $\{111\}$ layers along $\langle 112 \rangle$. We calculated the GSFE slip path and the twinning paths using standard methods~\cite{Vitek1968IntrinsicCrystals,Duesbery1998OverviewMetals,Tadmor2004ACrack,VanSwygenhoven2004StackingMetals}.
   
   Figures~\ref{fig:bain_sf}c and~\ref{fig:bain_sf}d show energies along the GSFE slip and twin paths, respectively. As before, we compare with seven EAM-based potentials. The ANI-Al potential agrees quite well with the reference DFT data for all measurements in Fig.~\ref{fig:bain_sf}. To quantify this agreement, we calculate the root mean squared error (RMSE), formed as an average over the Bain, Trigonal, GFSE slip, and GFSE twinning paths. ANI-Al achieves RMSE values of \unit{4.5}{meV/atom}, \unit{6.0}{meV/atom}, \unit{16.6}{mJ/m^2}, and \unit{11.4}{mJ/m^2}, respectively. For predicting these paths, the best classical potential is by Mishin et al.\cite{Mishin1999InteratomicCalculations}, which achieves errors of \unit{4.3}{meV/atom}, \unit{37.6}{meV/atom}, \unit{52.5}{mJ/m^2}, and \unit{15.9}{mJ/m^2}. 
   Figure~S2 of the SI quantifies the errors for each potential, averaged over the strain path. It is interesting to note that the Winey et al. potential\cite{Winey2009AAluminum}, which does exceptionally well in predicting many FCC properties (see Table~S3 of the SI), struggles to accurately predict the Bain and GSFE slip paths.

   Errors in modeling the BCC and FCC energy barriers can have severe consequences in MD simulations. We will show an example in Sec.~\ref{results:meltcurve}, where the Mendelev et al. potential\cite{Mendelev2008AnalysisCu} predicts transformation from FCC at BCC at just \unit{20}{GPa}, whereas the physically correct transition pressure should be hundreds of GPa.

  \subsection{Predicting radial distribution functions} 
  
  \begin{figure*}
    \includegraphics[width=\linewidth]{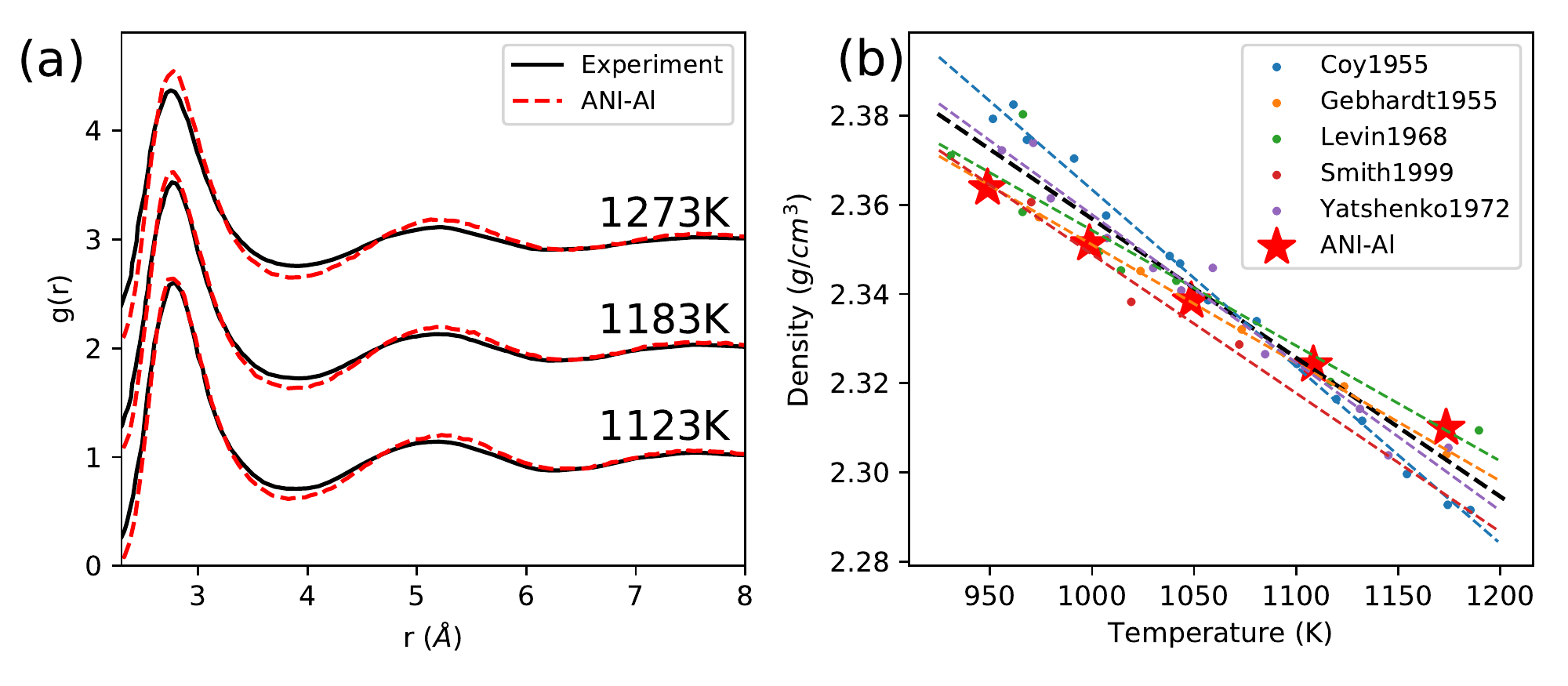}
    \caption{Molecular dynamics simulation in melt using the ANI-Al potential. (a) Radial distribution function at temperatures \unit{1123}{K}, \unit{1183}{K}, and \unit{1273}{K} compared to experiment~\cite{Mauro2011HighAl} (black line). (b) Density predictions as a function of temperature. The dashed black line is linear fit to all five sources of experimental data.}
    \label{fig:rdfdensity}
  \end{figure*}
  
  To validate our ANI-Al model in the liquid phase, we carry out MD simulations to measure radial distribution functions (RDF) and densities at various temperatures.
  Figure~\ref{fig:rdfdensity}a compares simulated RDFs with experimental measurements~\cite{Mauro2011HighAl} at \unit{1123}{K}, \unit{1183}{K}, and \unit{1273}{K}. Independent simulations were performed in the isobaric-isothermal ($NPT$) ensemble to determine equilibrium densities of liquid Al at the relevant ($P$,$T$) conditions. MD simulations of 2048 atoms were initialized at these densities and equilibrated for \unit{50}{ps} in the $NVT$ ensemble using the Nosé-Hoover-style equations of motion~\cite{Hoover1985CanonicalDistributions} derived by Shinoda et al~\cite{Shinoda2004RapidStress}. Reported RDFs were calculated (bin size of \unit{0.05}{\angstrom}) by averaging 100 instantaneous RDFs, which were \unit{0.1}{ps} apart, in the final \unit{10}{ps} of the $NVT$ equilibration. A timestep of \unit{1}{fs} was used for these simulations. Figure~\ref{fig:rdfdensity}b compares ANI-Al predicted densities at various temperatures (still at atmospheric pressure) to multiple experimental values~\cite{Assael2006ReferenceIron,Smith1999MeasurementTechnique,YatsenkoSPKononenkoVI1972ExperimentalGa,CoyW.J.Mateer1955CoyMateer,Levin1968LevinDensity,GebhardtE.BeckerM.andDorner1955GebhardtAl}. For reference, the melting temperature is $T_\textrm{melt} \approx \unitm{933}{K}$. The agreement between ANI-Al predictions and experiment is comparable to the variation between different experiments.
  
  \subsection{Predicting FCC phonon spectrum} 
    \label{results:phonon}
\begin{figure}
    \includegraphics[width=\linewidth]{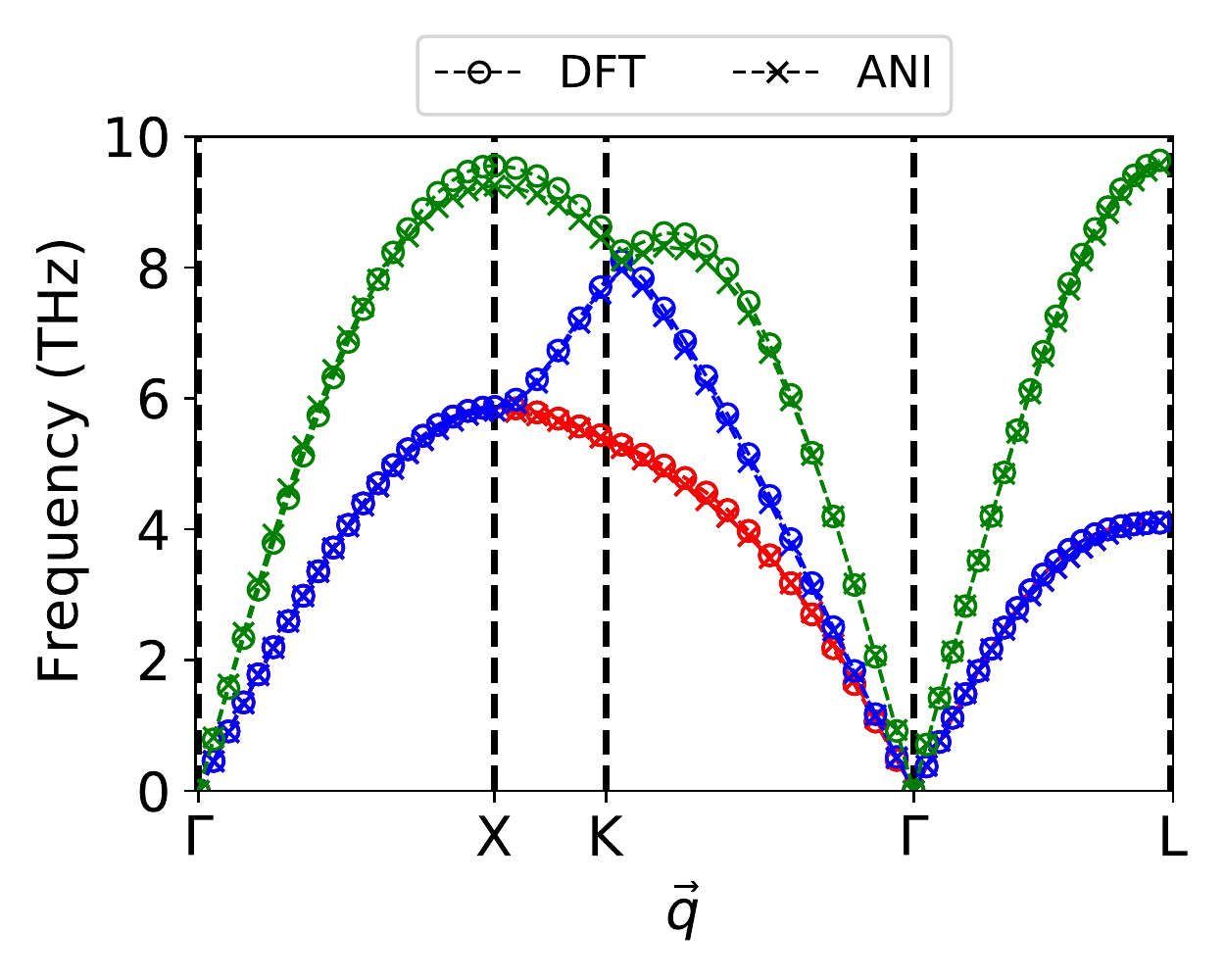}
    \caption{Phonon spectrum of FCC Al predicted by the ANI-Al model compared to DFT. }
    \label{fig:phonon_spectrum}
  \end{figure}

Figure~\ref{fig:phonon_spectrum} compares the ANI-Al predicted phonon spectrum to that of DFT. In both cases, the frequencies were calculated using the PHON program~\cite{Alfe2009PHON:Method} via the small-displacement method~\cite{Kresse1995AbGraphite,Alfe2001ThermodynamicsConditions}. A super-cell of size $4\times 4\times 4$ FCC unit cells was used for the calculations. The ion at the origin of this super-cell was displaced in [100], with a magnitude of 1\% the equilibrium FCC lattice spacing, and the forces were calculated on all the ions. These forces were used to calculate the phonon frequencies in the quasi-harmonic approximation. Figure~\ref{fig:phonon_spectrum} shows good agreement between ANI and DFT predictions of FCC Al phonon spectrum.

\subsection{Predicting FCC point defects} 
    \label{results:ptdefects}

ANI-Al predicts the formation energies for vacancy and ($\langle 100 \rangle$ dumbbell) interstitial defects to be \unit{663}{meV} and \unit{2.49}{eV}, respectively. The corresponding DFT predictions are \unit{618}{meV} and \unit{2.85}{eV}. The vacancy formation energy is experimentally estimated to be about \unit{680}{meV}~\cite{Schaefer1987PositronAluminium}. Tables S3 and S4 of the SI also list predictions for existing classical and ML potentials. The relatively large deviation between ANI-Al and DFT predictions is perhaps an indication that vacancies and interstitials did not play a large role in the configurations sampled during the active learning procedure.

\subsection{Predicting FCC surface properties} 
    \label{results:surfaces}
The properties of surfaces predicted by our ANI-Al model is compared to values from DFT, experiments, and seven EAM-based potentials in Table S3 of the SI. Table S4 of the SI compares with previous ML results, where available. ANI slightly over-predicts the surface energy for \{100\}, \{010\}, and \{111\}, with a maximum error of 6.6\% (for \{100\}) compared to DFT predictions. ANI predicts the correct sign for surface relaxation (inward or outward) in all but one case ($d_{12}^{\{100\}}$). The outward relaxation of \{100\} and \{111\} surfaces in Al are considered ``anomalous'' and ANI predicts this correctly only for \{111\}, despite correct predictions by DFT for both surfaces. Also note that ANI-Al correctly predicts the ordering of the magnitudes of surface relaxation, $|d_{12}^{\{110\}}| > |d_{12}^{\{100\}}|  \approx |d_{12}^{\{111\}}| $, but the quantitative agreement with DFT reference calculations is poor. The ANI-Al training dataset includes only bulk systems with periodic boundary conditions, but some surface configurations may have been incidentally sampled due to void formation at low densities.
    
  \subsection{Predicting liquid-solid phase boundaries} 
    \label{results:meltcurve}

  \begin{figure}
    \includegraphics[width=\linewidth]{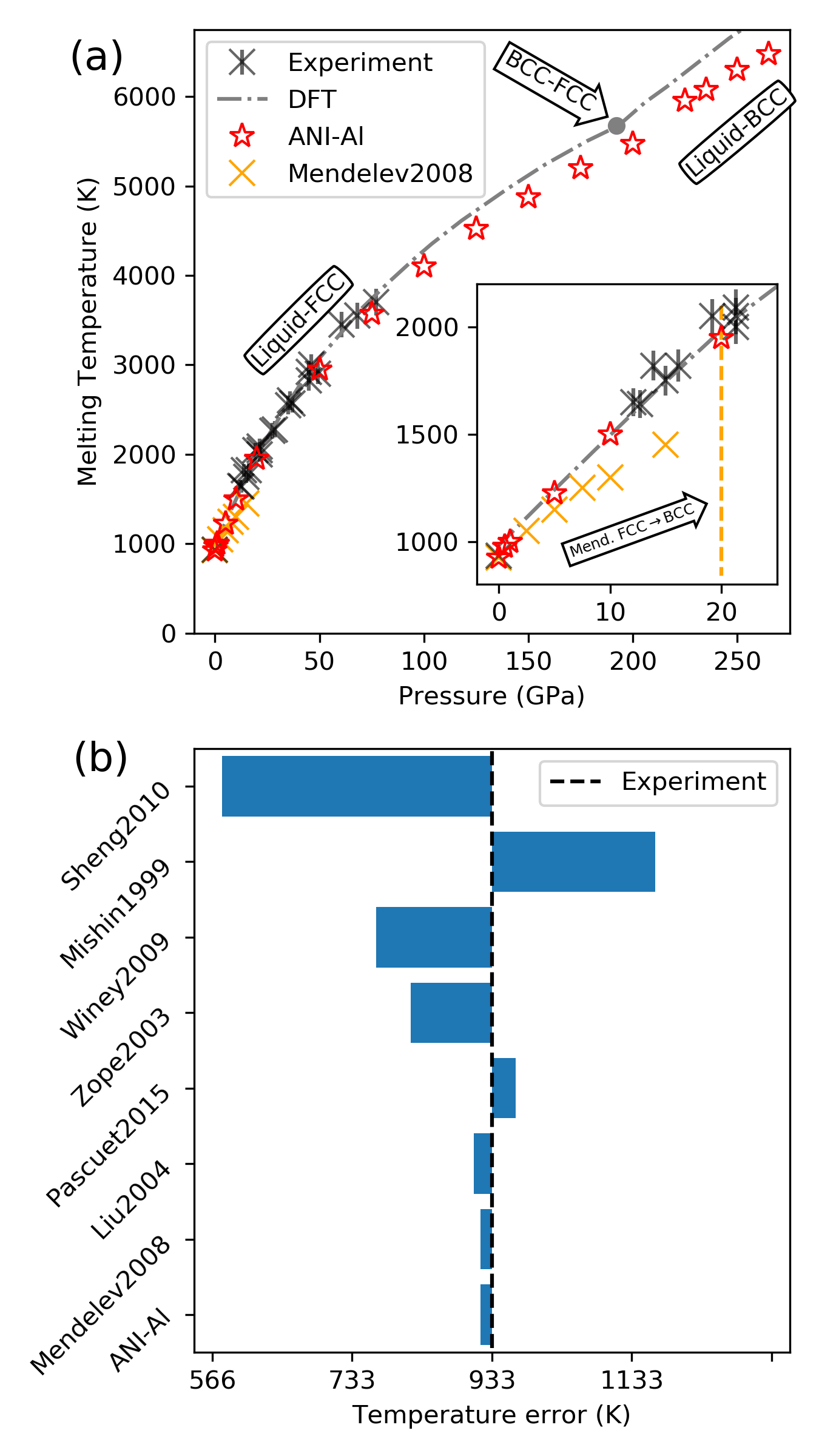}
    \caption{(a) Aluminum melt curves calculated from DFT~\cite{Sjostrom2016MultiphaseTheory}, ANI-Al, and the Mendelev et al. EAM potential~\cite{Mendelev2008AnalysisCu}, compared with experimental data~\cite{Hanstrom2000HighAluminium}. Below \unit{210}{GPa} we show FCC-liquid coexistence. Above \unit{210}{GPa} we show BCC-liquid coexistence. The inset zooms to pressures from 0 to \unit{20}{GPa}. (b) Errors in predicting the melt temperature at atmospheric pressure.}
    \label{fig:melt_curve}
  \end{figure}
  
    Figure~\ref{fig:melt_curve} shows the liquid-solid coexistence line in the pressure-temperature plane. At each pressure, we calculated the coexistence temperature by performing simulations with an explicit solid-liquid interface~\cite{Morris1994MeltingPhases,Morris2002TheSimulations,Espinosa2013OnModel}. The details of these simulations are provided in Sec. 1.7.1 of the SI. Experimental data is available up to about \unit{100}{GPa}~\cite{Hanstrom2000HighAluminium}. We also compare with prior DFT calculations~\cite{Sjostrom2016MultiphaseTheory} and a classical MD potential. For the latter, we used the Mendelev et al. potential~\cite{Mendelev2008AnalysisCu}, which was explicitly parameterized to model the melting point of aluminum, $T_\textrm{melt} \approx \unitm{933}{K}$ at atmospheric pressure. At this pressure, both Mendelev and ANI-Al potentials predict an FCC melting point of about \unit{925}{K}, in good agreement with experiment.
    
    The Mendelev model begins to underestimate the melting temperature at around \unit{5}{GPa}, whereas the ANI-Al model remains quite accurate up to about \unit{50}{GPa}. Note that the ANI-Al training data was restricted to a limited range of densities (yellow region of Fig.~\ref{fig:coldcurveAl}a) which correspond to pressures up to about \unit{50}{GPa} (Fig.~S1 of the SI). We were surprised to observe qualitative agreement between the ANI-Al and DFT predicted coexistence curves up to \unit{250}{GPa}, even though this is a significant extrapolation for ANI-Al.
    
    For the Mendelev simulations, the liquid-FCC coexistence curve only extends to about \unit{20}{GPa}; beyond that point we observed nucleation into BCC.
    According to prior DFT-based studies~\cite{Tambe2008BulkStudy,Sjostrom2016MultiphaseTheory}, and experiment~\cite{Fiquet2019StructuralGPa}, the solid-to-solid transition out of FCC should require hundreds of GPa. Figure~\ref{fig:melt_curve} includes the theoretically predicted liquid-BCC coexistence curve at pressures between 225 to \unit{275}{GPa}.
    
 \subsection{Phase transition dynamics} \label{results:phasechange}
   
     \begin{figure*}
   \centering
    \includegraphics[width=1.0\textwidth]{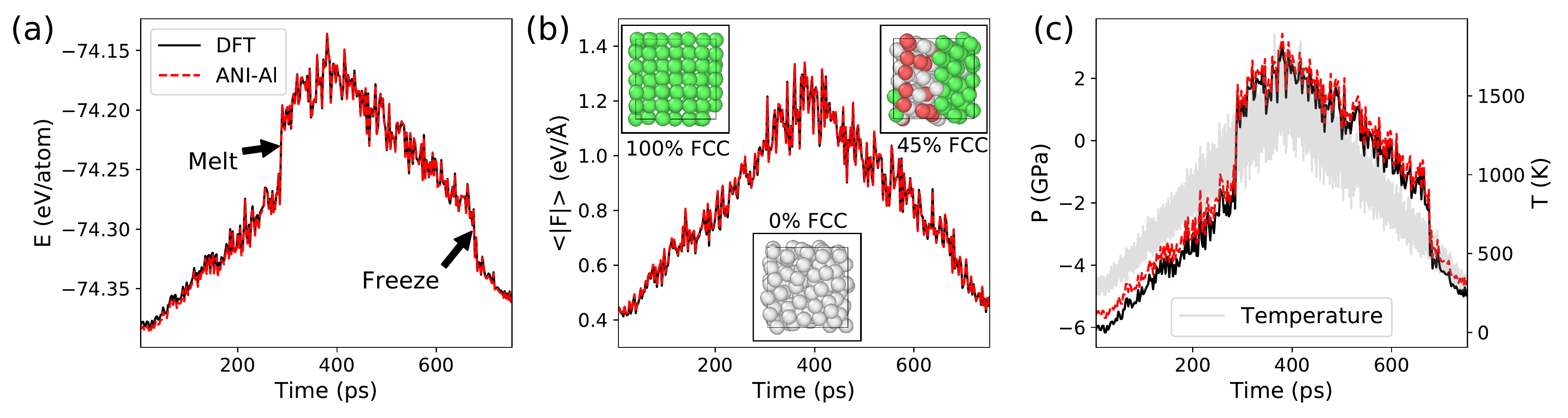}
    \caption{ML-driven molecular dynamics, showing melting and freezing processes. The system is heated from \unit{300}{K} to \unit{1500}{K}, and cooled back to \unit{300}{K}. Reference DFT calculations (black) are used to verify the ANI-Al predictions (red) for the energy, mean (avg.) force, and pressure. The instantaneous temperature is shown in grey on the rightmost panel. The middle panel insets show the local atomic structure (green--FCC; gray--disordered; red--HCP) at snapshots before melting, after melting, and after refreezing.}
    \label{fig:almeltfreeze}
   \end{figure*}
   
   Next we carry out a nonequilibrium MD simulation to observe both freezing and melting dynamics. Our intent is to verify the ANI-Al predicted energies and forces at snapshots along the dynamical trajectory. Along the trajectory the temperature is slowly increased from \unit{300}{K} to \unit{1500}{K}, then cooled back to \unit{300}{K}. The details of these simulations are provided in Sec.~1.7.2 of the SI.
   
   Figure~\ref{fig:almeltfreeze} shows the potential energy, mean force magnitude, and pressure for both ANI-Al and DFT along this trajectory. Melting from FCC to liquid occurs at around \unit{300}{ps} and freezing occurs around \unit{700}{ps}. Pressure was calculated using the method of Ref.~\onlinecite{Thompson2009GeneralConditions}.
   The inset images in the middle panel of Figure~\ref{fig:almeltfreeze} show the composition of the system before and after melting, and after refreezing. Compositional information was obtained using the Common Neighbor Analysis as implemented in the OVITO visualization software~\cite{Stukowski2010VisualizationTool}.
   
   Every \unit{2.5}{ps} along the trajectory we sampled a frame to perform reference DFT calculations. The error between ANI-Al and DFT is generally small. Averaged over the full trajectory, the MAE for energy is  \unit{0.84}{meV/atom}. The MAE for each force component individually is \unit{0.023}{eV/\angstrom}. The MAE for ANI-Al predicted pressure is \unit{0.36}{GPa}. Interestingly, there is a systematic tendency for ANI-Al to overestimate pressure, especially at negative pressures. This seems a bit surprising, because the ANI-Al force predictions seem reasonably good, and these determine pressure through the atomic virial tensor. Perhaps the tendency to ANI-Al overestimate pressure is a reflection the fact that a large fraction of its training data was sampled at very large positive pressures (cf. Fig.~S1 in the SI).
  
   Figures~S3 and~S4 of the SI further verify the ANI-Al force predictions for MD simulations over large a range of temperatures and densities.
  
\subsection{Simulation of shock physics} \label{results:shockdynamics}
Finally, to verify our potential at predicting material response under extreme conditions, we carried out a large scale shock simulation using NeuroChem interfaced to the LAMMPS molecular dynamics software package~\cite{Plimpton1995FastDynamics}. The simulation cell, containing about 1.3M atoms, has approximate dimensions $\unitm{10}{nm} \times \unitm{211}{nm} \times \unitm{10}{nm}$ in the $x=[112]$, $y=[\bar{1}10]$, and $z=[\bar{1}\bar{1}1]$ crystallographic directions. Prior to shock, the volume was equilibrated at \unit{300}{K} for \unit{15}{ps} in the $NVT$ ensemble. Periodic boundary conditions were applied in $x$ and $z$, with free-surfaces in $y$. After equilibration, a $u_p = \unitm{1.5}{km/s}$  shock was applied in $y$ using the reverse-ballistic configuration~\cite{Frohlich2014MolecularResponses} and the system was evolved in the $NVE$ ensemble. In this method, a rigid piston is defined by freezing a rectangular block of atoms and the velocities of remaining atoms are modified by adding $-u_p$ to the $y$-component. This sets up a supported shock wave in the flexible region of the simulation cell which was evolved for \unit{50}{ps}. Using spatial domain decomposition as implemented in LAMMPS, the 1.3M atoms were distributed across 80 Nvidia Titan V GPUs, and the required wall-clock time for the entire \unit{65}{ps} simulation (65k MD timesteps) was about 15 hours.

\begin{figure*}
    \includegraphics[width=\linewidth]{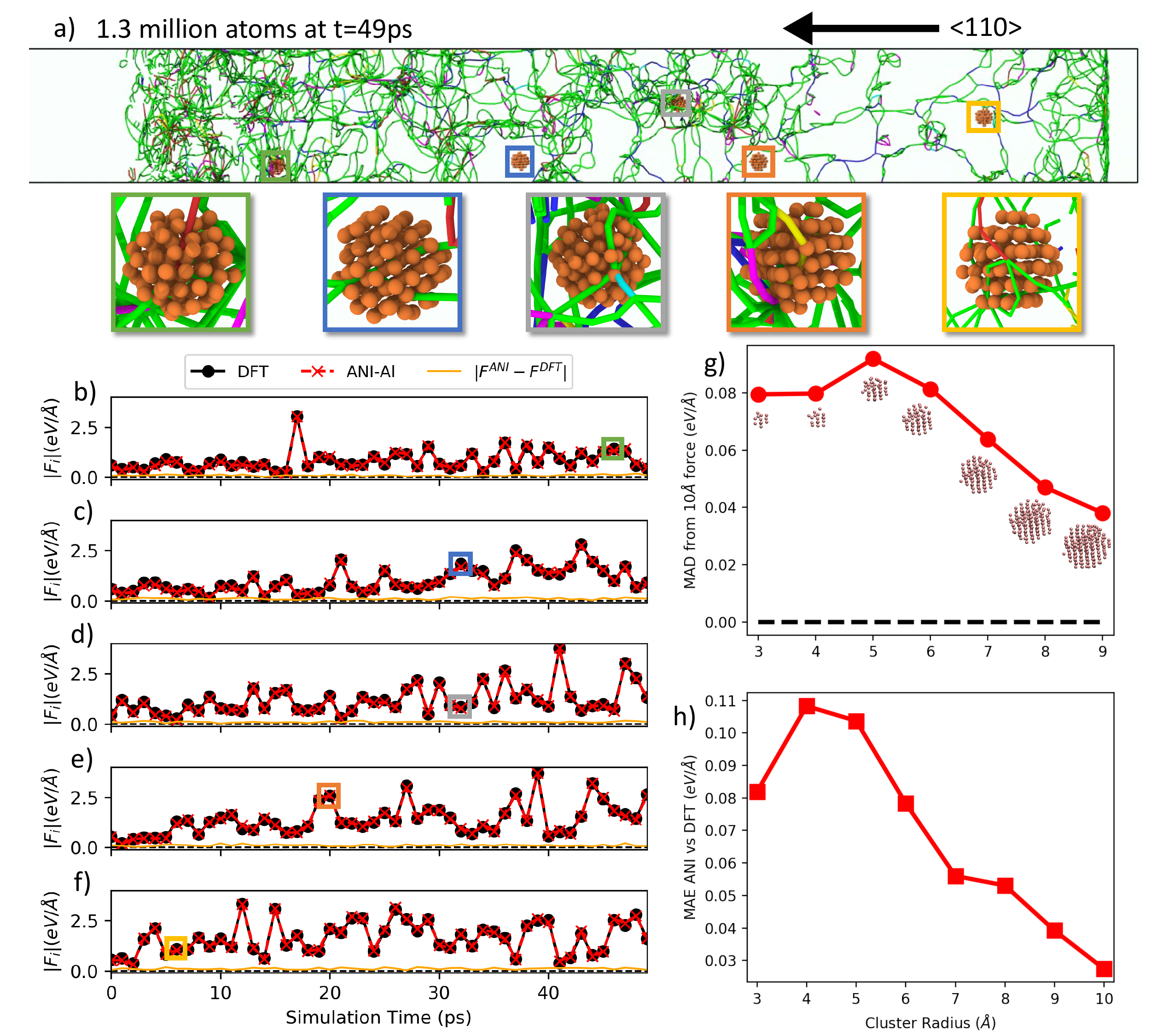}
    \caption{A 1.3 million atom shock simulation using the ANI-Al potential. A shock of 1.5 km/s was initiated from the right along the $\langle 110\rangle$ crystallographic direction. (a) Dislocation structure after \unit{49}{ps} of simulation as well as zooms for five randomly selected atoms at hand-picked times. (b)-(f) Verification of the ANI-Al force predictions for these five atoms every \unit{1}{ps}. Reference forces were obtained by performing new DFT calculations for small clusters centered the five atoms. (g) Comparison of DFT-calculated forces on the central atom for varying cluster radius (reference force calculated at radius \unit{10}{\angstrom}). (h) Mean absolute error of ANI-Al predicted forces, relative to DFT, as a function of cluster radius.}.
    \label{fig:shock}
\end{figure*}

Figure~\ref{fig:shock}a shows the dislocation structure in the simulation cell (as predicted by the Dislocation Extraction Algorithm (DXA)\cite{Stukowski2010VisualizationTool,Stukowski2014AModels} at \unit{49}{ps}.

We randomly selected five atoms in the simulation volume for further analysis. The atomic environments for these five atoms are shown as clusters and highlighted with colored boxes in Fig.~\ref{fig:shock}a. The five zoomed insets illustrate that dislocations can pass near each of the five central atoms at specific times, which are marked with colored boxes in Figs.~\ref{fig:shock}b-f.

Figures~\ref{fig:shock}b-f compare the ANI-Al predicted forces with new reference DFT calculations at every \unit{1}{ps} of simulation time. For each sample point, a local environment (a cluster of radius \unit{7}{\angstrom}) was extracted from the large scale shock simulation and placed in vacuum. A new DFT calculation was performed on this cluster, and the resulting force on the central atom was compared to the corresponding ANI-Al prediction. As shown in Figs.~\ref{fig:shock}b-f, the magnitudes of the forces have a characteristic scale of order \unit{1}{eV/\angstrom}. The mean absolute error, for the ANI-Al predictions of each force component individually, is approximately \unit{0.06}{eV/\angstrom}. However, as we will discuss below, artificial surface effects due to finite cluster radius $r=\unitm{7}{\angstrom}$ cannot be neglected, and larger clusters are required to measure the true ANI-Al error for these shock simulations.

To systematically study the effect of cluster cutoff radius $r$, we further down-sampled to 10 local atomic environments. Figure~\ref{fig:shock}g quantifies the $r$-dependence on the DFT-calculated force $\mathbf f_{r}$. Specifically, it shows the mean of $|f_{r; a} - f_{r_0; a}|$, where the reference radius is taken to be $r_0 = \unitm{10}{\angstrom}$. Averages were taken over all force components $a=x,y,z$ and over all 10 local atomic environments.
Surface effects for $r = \unitm{7}{\angstrom}$ are seen to modify the central atom force by about \unit{0.06}{eV/\angstrom}, which is of the same order as the ANI-Al disagreement with DFT, when measured using this $r$. The average force magnitude for these 10 configuration samples is \unit{1.12}{eV/\angstrom}, so the observed deviations at $r = \unitm{7}{\angstrom}$ represent about a 5\% effect.

Figure~\ref{fig:shock}h illustrates that ANI-Al and DFT agreement becomes better for calculations on larger clusters, i.e., where artificial surface effects are reduced. With cluster radius $r=\unitm{7}{\angstrom}$, the ANI-Al mean absolute error (MAE) for force components is about \unit{0.06}{eV/\angstrom}. At the largest cluster size we could reach ($r = \unitm{10}{\angstrom}$) the ANI-Al MAE reduces to about \unit{0.03}{eV/\angstrom}, i.e., about a 3\% relative error. For reference, recall that in the liquid-solid phase transition simulations of Sec.~\ref{results:phasechange}, the average ANI-Al force errors were slightly lower, at \unit{0.023}{eV/\angstrom}; in that context, however, the reference DFT calculations did not suffer from artificial surface effects.

It makes sense that ANI-Al and DFT forces are most consistent for the largest cluster sizes, given that the training data produced by active learning consists entirely of bulk systems.
Note that although the nominal ANI-Al cutoff radius is just \unit{7}{\angstrom}, the model can still generate strong effective couplings at distances of up to \unit{10}{\angstrom} through intermediary atoms that process angular features, as described in Sec.~1.4 of the SI.
   
  \section{Importance of active learning}
  In this section we explore the advantages of our automated active learning approach, which minimizes reliance on human knowledge. We first compare with a na{\"i}ve approach for data generation via a near-equilibrium MD simulation, and second characterize the chemical space covered by our automated sampling methodology.
  
  \subsection{Human Knowledge vs. Automated Sampling}
  The success of ANI-Al hinges on the diversity of the active learned dataset. To demonstrate this, we compare ANI-Al against an ML model trained on a much more limited dataset. We will call this baseline dataset ``FCC/Melt,'' as it consists only of samples from the FCC and liquid phases. Specifically, the FCC/Melt dataset is constructed by taking regular snapshots from near-equilibrium MD trajectories. For each snapshot, we perform a DFT calculation to determine the reference energy and forces.
  
  The first such MD trajectory is shown in Fig.~\ref{fig:almeltfreeze}. There, 108 atoms were initialized to FCC, heated from \unit{300}{K} to \unit{1500}{K}, and cooled back to \unit{300}{K}. We take 300 snapshots from this trajectory, equally spaced in time, to add to the FCC/Melt dataset. For increased variety, the FCC/Melt dataset contains an additional 250 DFT calculations taken from the liquid phase over a range of temperatures and pressures (Sec.~1.7.3 in the SI contains details). In sum, the FCC/Melt dataset contains 550 DFT calculations for near-equilibrium FCC and liquid configurations.
    
  \begin{table}
    \caption{Performance of ANI models trained on active learning (AL) and near-equilibrium FCC/Melt datasets. We compare MAE/RMSE values for held out test data from AL and FCC/Melt datasets.}

    \centering 
    
    \begingroup
    \setlength{\tabcolsep}{10pt} 
    \begin{tabular}{l l l} 
    \hline\hline 
    Model type & FCC/Melt test & AL test \\  
    \hline
    \multicolumn{1}{c}{} & \multicolumn{2}{c}{Energy error (meV/atom)} \\
    \hline 
    FCC/Melt trained  & 2.0/4.0   & 40/110 \Tstrut\\
    AL trained & 1.4/1.9   & 1.3/1.9  \\ 
    \hline
             & \multicolumn{2}{c}{Force component error (eV/\AA)} \\
    \hline
    FCC/Melt trained & 0.04/0.07 & 0.49/1.53 \\
    AL trained & 0.03/0.04 & 0.04/0.06  \\ [1ex] 
    \hline 
    \end{tabular}
    \endgroup
    
    \label{table:disordervsfccmelt} 
  \end{table}
  
  Table~\ref{table:disordervsfccmelt} compares our ANI-Al model, trained on the full active learned (AL) dataset, to an ANI model trained on the much more restricted FCC/Melt dataset. The two model types are compared by testing on held out portions of both datasets. Figures~S5 and~S6 in the SI show the associated correlation plots.
  
  A conclusion of Table~\ref{table:disordervsfccmelt} is that both the AL trained and FCC/Melt trained models have comparable errors when predicting on the held out FCC/Melt test data. However, when testing on the held out AL data, the FCC/Melt trained model does quite poorly. This failure casts doubt on the ability of the FCC/Melt trained model to study new dynamical physical processes: Will a rare event occur that pushes the FCC/Melt trained model outside its range of validity? To mitigate this danger it is essential to make the training dataset as broad as possible, which is our aim with active learning.
  
  \begin{figure*}
    \includegraphics[width=\linewidth]{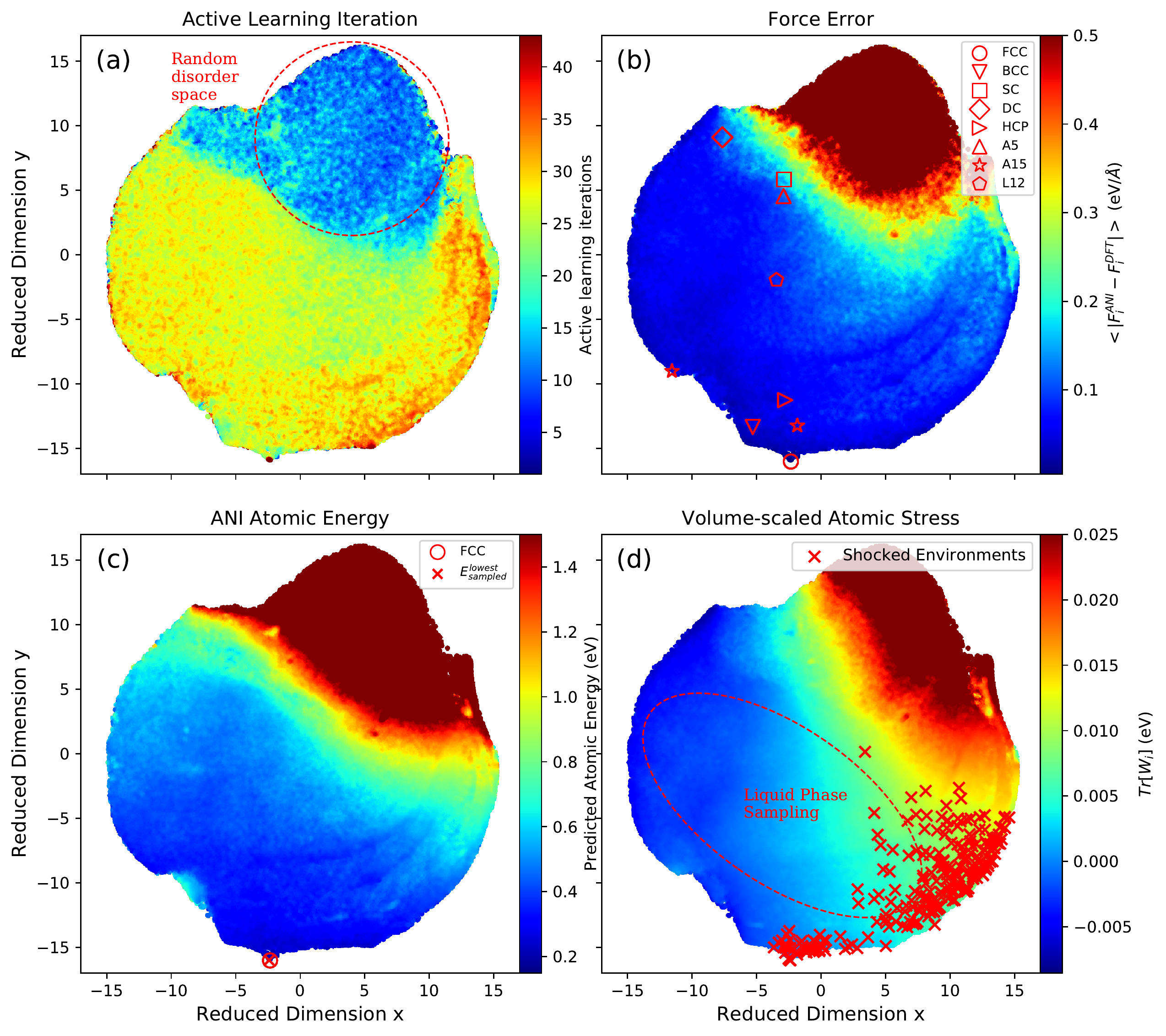}
    \caption{Visualization of the atomic environments contained within the active learned training dataset, using the t-distributed Stochastic Neighbor Embedding (t-SNE) method. Radial neighbor regression (RNR) is is used to color the average property within a radius of a given point in the 2D embedding space. \textbf{(a)} Active learning iteration at which a sample was taken; disordered space is circled. \textbf{(b)} Force error; eight crystal structures are marked. \textbf{(c)} ANI predicted atomic energy; FCC is observed to be the lowest energy configuration in the embedding space. \textbf{(d)} Volume-scaled atomic stress; shocked environments are marked and liquid environments are circled.}
    \label{fig:TSNE}
  \end{figure*}
  
  \subsection{Coverage of Chemical Space}
  In this section we characterize the sampling space covered by our active learning methodology using the t-distributed Stochastic Neighbor Embedding~\cite{Hinton2008VisualizingT-SNE} (t-SNE) method as implemented in the OpenTSNE~\cite{Policar2019OpenTSNE:Embedding} Python package. In Figs.~\ref{fig:TSNE}a-d, every local atomic environment in the active learned training dataset is mapped onto a reduced, two-dimensional space.  Hyper-parameters of the t-SNE embedding process are presented in Sec.~1.6 of the SI. In brief, we use the activations after the first layer of the ANI-Al neural network as an abstract representation (``latent space vector'') of the \unit{7}{\angstrom}-radius local atomic environment around each atom. The cosine distances between all pairs of these latent space vectors (for all points of the dataset) are the inputs to t-SNE. The output of t-SNE is, ideally, a mapping of all latent space vectors onto the two dimensional embedding space that, in some sense, is maximally faithful to pairwise distances. t-SNE thus provides a two-dimensional visualization of all atoms in all configurations of the dataset.
  
  Figures~\ref{fig:TSNE}a-d use radial neighbor regression (RNR) to associate atomic environments (averaged within the embedding space) with four different properties. Figure~\ref{fig:TSNE}a shows average active learning iteration count, Fig.~\ref{fig:TSNE}b shows the average force error (saturated at \unit{0.5}{eV/\angstrom}), Fig.~\ref{fig:TSNE}c shows the ANI predicted atomic energy (saturated at \unit{1.5}{eV}), and Fig.~\ref{fig:TSNE}d shows the trace of the ANI-Al predicted atomic stress tensor (saturated at \unit{0.025}{eV/\angstrom}).
  
  Observe that the sampled points are well connected in the reduced dimensional space, and not clustered. In contrast, a poorly sampled dataset would typically lead to obvious clusters, clearly separated by gaps. In Fig~\ref{fig:TSNE}a one sees that active learning procedure progresses from sampling random disordered configurations (blue region at top) to sampling much more structured data. The left, bottom and right edges of the embedding space was not sampled until late in the active learning process (red). Up until about about 10 iterations into the active learning procedure, all MD sampling trajectories never ran long enough to make it to an ordered atomic configuration (recall that the trajectories end once they reach a configuration with very high ML uncertainty).
  Despite being very well sampled, comparison with Fig.~\ref{fig:TSNE}b shows that the ML model still has greatest difficulty in fitting this disordered (high entropy) region of configuration space.
  Figures~\ref{fig:TSNE}c and \ref{fig:TSNE}d show that these disordered atomic environments typically have high energies and stresses. 
  
  
  Markers in Fig.~\ref{fig:TSNE}b show the local atomic environments for perfect crystals; we selected eight crystal structures that could potentially compete with FCC as the ground state.  Observe that all eight markers lie within the sampled space (interestingly, only FCC and A15 crystals are placed at the edge), and are continuously connected. The average force error in the region of all crystal structures is generally very low (less than \unit{0.1}{eV/\angstrom}), except for the simple cubic and diamond cubic regions, which are very high-energy structures, and thus less physical. Figure~\ref{fig:TSNE}c shows that the position of FCC is almost perfectly overlapping the lowest energy configuration sampled during active learning. As mentioned above, the FCC structure was not found until at least 10 active learning iterations. Later in the active learning process, however, local FCC configurations became quite well sampled (cf. Fig.~\ref{fig:TSNE}a).
  
  Figure~\ref{fig:TSNE}d illustrates with red crosses the shocked environments randomly sampled from the simulations of Sec.~\ref{results:shockdynamics}. Interestingly, these samples are largely confined to the bottom right portion of embedding space, and span a fairly significant range of local atomic stresses. 
  Early in the shock simulation, the atomic environments live primarily near the FCC region of the embedding space, with small local stress. As the shock wave passes through each local environment, one can sample much higher pressure and temperatures conditions. Afterwards, there remains a complicated pattern of defects. Importantly, throughout the entire shock process, all visited atomic environments appear to be well represented by the training dataset. This is consistent with the fact that the force errors of Fig.~\ref{fig:TSNE}b appear to be remain small for all regions (e.g., bottom-edge of embedding space) where the shocked environments appear. The region circled and labeled "Liquid phase sampling" was obtained from the atomic environments in the liquid phase simulations shown in Fig.~S4 and described in Sec.~S1.7.3 of the SI. The configurations appearing in a shock are largely distinct from those appearing in simulations of the liquid phase.
  
 \section{Outlook}
 
 ML is emerging as a powerful tool for producing interatomic potentials with unprecedented accuracy; recent models routinely achieve errors of just a couple meV per atom, as benchmarked over a wide variety of ordered and disordered atomic configurations. Here, we presented a technique to automatically construct general purpose ML potentials that requires almost no expert knowledge.
 
 Modern ML potentials can be used for large-scale MD simulations. To quantify performance, consider for example the optimized Neurochem code~\cite{2017NeuroChemgithub.com/atomistic-ml/neurochem,Smith2017} applied to ANI-Al with an 8x ensemble of neural networks, and a simulation volume of thousands of atoms; here, we measure up to 67k atom time-steps per second when running on a single Nvidia V100 GPU. With 80 GPUs, our current LAMMPS interface (not fully optimized) achieved 1.6M atom time-steps per second for the 1.3M-atom shock simulation. A study conducted parallel to ours performed ML-MD simulations of 113M copper atoms by using 43\% of the Summit supercomputer (about 27k V100 GPUs)~\cite{Jia2020PushingLearning}. The speed of ANI-Al is perhaps two orders of magnitude slower than an optimized EAM implementation, but vastly faster than \textit{ab initio} MD would be.
  
  Because ML models are so flexible, the quality and diversity of the training dataset is crucial to model accuracy. Here we focused on the task of dataset construction and, specifically, sought to push the limits of \emph{active learning}. We presented an automated procedure for building ML potentials. The required inputs are limited to: physical parameters such as the temperature and density ranges over which to sample, the interaction cutoff radius for the potential (we selected \unit{7}{\angstrom} for aluminum), and some ML hyperparameters that were largely reused from previous studies. We did \emph{not} include any expert knowledge about candidate crystal ground states, defect structures, etc. Nonetheless, the active learning procedure eventually collected sufficient data to produce a broadly accurate potential for aluminum.
  
  We emphasize that the starting point for the active learning procedure consisted of DFT calculations for completely disordered configurations. As the ML potential improved, the quality of the MD sampling increased, and the training data collected could become more physically relevant. The timeline of this ``discovery'' process is illustrated in Fig.~\ref{fig:TSNE}a. After about 10 active learning iterations (1000+ DFT calculations), the ML potential became robust enough that the MD simulations could nucleate crystal structures. From this point onward, the ML predictions for crystal properties could rapidly improve.
  
  Previous potential development efforts have benefited from careful dataset design. Our decision to pursue a fully automated approach certainly made the modeling task more difficult, but was motivated by our belief that defects appearing in real, highly-nonequilibrium processes may be difficult to fully characterize \emph{a priori}. As an example, consider the complicated dislocation patterns appearing in the shock simulation of Fig.~\ref{fig:shock}. It would likely be difficult to hand-design a dataset that fully captures all defect patterns appearing in shock. Active learning, however, seems to do a good job of sampling the relevant configuration space (cf. the marked points in Fig.~\ref{fig:TSNE}d). Indeed, throughout the entire shock simulation, the ANI-Al predicted forces are in good agreement with new reference DFT calculations, even for atoms very near dislocation cores. Even though most of the active learned training data is far from perfect FCC, the ability of ANI-Al to predict aluminum FCC properties seems roughly in line with other recent ML studies, as shown in Table~S4 of the SI~\cite{Zhang2019ActiveSimulation,Pun2019PhysicallyinformedMaterials}.
  
  A challenge for the active learning procedure presented in this work is its large demand on computational resources. Our final active learned dataset contains over 6,000 DFT calculations; each calculation was performed on a supercell containing up to 250 atoms. For future work, it would be interesting to explore whether the majority of the training data could be weighted toward much smaller supercells. It would also be interesting to investigate ways to make the active learning more efficient, e.g., by systematically studying the effect of various parameters required by the procedure. Other areas for improvement may include: employing a dynamics with modulated stress or strain, smarter sampling that goes beyond nonequilibrium MD~\cite{Karabin2020AnPotentials}, and better estimation of the ML error bars.
  
\acknowledgments

This work was partially supported by the LANL Laboratory Directed Research and Development (LDRD) and the Advanced Simulation and Computing Program (ASC) programs. Within ASC, we acknowledge support from the Physics and Engineering Modeling (ASC-PEM) subprogram and the Advanced Technology Development and Mitigation (ASC-ATDM) subprogram. We acknowledge computer time on the Sierra supercomputing cluster at LLNL, Institutional Computing at LANL, and the CCS-7 Darwin cluster at LANL. JSS was supported by the Nicholas C. Metropolis Postdoctoral Fellowship. NM and JSS were partially supported by the Center for Nonlinear Studies (CNLS). This work was performed, in part, at the Center for Integrated Nanotechnologies, an Office of Science User Facility operated for the U.S. Department of Energy (DOE) Office of Science.

\section*{Data availability}

The active learned training dataset and final ANI-Al potential will be released at \url{https://github.com/atomistic-ml/ani-al}.

\section*{Code availability}

Two implementations of the ANI neural network architecture are available online: TorchANI (\url{https://github.com/aiqm/torchani}) and NeuroChem (\url{https://github.com/atomistic-ml/neurochem}).

\bibliographystyle{apsrev4-1}
\bibliography{references-1}

\end{document}